\documentclass[12pt,preprint]{aastex}

\shorttitle{Properties of Dwarf Galaxies, Virgo Sample}
\shortauthors{O. Vaduvescu, M. L. McCall, M. G. Richer}
\slugcomment{Accepted by AJ. Full version including Figure 1 at \\
             http://aries.phys.yorku.ca/$\thicksim$ovidiuv/Paper3.ps.gz (5 MB)}

\begin{document}

\title{Infrared Properties of Star Forming Dwarf Galaxies: \\
       II. Blue Compact Dwarf Galaxies in the Virgo Cluster\footnote{These 
       data were acquired at the Observatorio Astronomico Nacional in San Pedro Martir, Mexico.}}

\author{Vaduvescu, Ovidiu}
\affil{York University, Department of Physics and Astronomy \\
       4700 Keele Street, M3J~1P3, Toronto, ON, Canada \\
       email: ovidiuv@yorku.ca}

\author{Richer, Michael G.}
\affil{
Observatorio Astronomico Nacional, Instituto de Astronomia, UNAM, \\ 
       PO Box 439027, San Diego, CA 92143-9027, USA \\
       email: richer@astrosen.unam.mx}  

\author{McCall, Marshall L.}
\affil{York University, Department of Physics and Astronomy \\
       4700 Keele Street, M3J~1P3, Toronto, ON, Canada \\
       email: mccall@yorku.ca}

\begin{abstract}
A sample of 16 blue compact dwarf galaxies (BCDs) in the Virgo Cluster has been imaged 
in the near-infrared (NIR) in $J$ and $K_s$ on the 2.1m telescope at OAN-SPM in Mexico. 
Isophotes as faint as $\mu_J$ = 24 mag arcsec$^{-2}$ and $\mu_{K_s}$ = 23 mag arcsec$^{-2}$ 
have been reached in most of the targets. 
\noindent
Surface brightness profiles can be fitted across the whole range of radii by the sum of 
two components: a hyperbolic secant (sech) function, which is known to fit the light 
profiles of dIs, and a Gaussian component, which quantifies the starburst near the centre. 
\noindent
Isophotal and total fitted NIR magnitudes have been calculated, along with semimajor axes at 
$\mu_J$ = 23 mag arcsec$^{-2}$ and $\mu_{K_s}$ = 22 mag arcsec$^{-2}$. The diffuse underlying 
component and the young starburst have been quantified using the profile fitting. Most color 
profiles show a constant color, between $J-K_s=0.7$ to $0.9$ mag. 
\noindent
The diffuse component represents the overwhelming majority of the NIR light for most BCDs, 
the starburst enhancing the flux by less than about 0.3 mag. 
\noindent
Linear correlations were found between the sech scale length and the sech magnitude, and between 
the sech semimajor axis and the sech magnitude. Overall, galaxies with more luminous diffuse 
components are larger and brighter in the centre. The central burst correlates with the diffuse 
component, with brighter BCDs having stronger star-bursts, suggesting that more massive objects 
are forming stars more efficiently. 
\noindent
BCDs lie on the ``fundamental plane'' defined by dwarf irregulars (dISs) in Paper I, following 
the same relation between sech absolute magnitude, sech central surface brightness, and the 
hydrogen line-width $W_{20}$, although the scatter is larger than for the dIs. On the other hand, 
correlations between the sech absolute magnitude and the sech central surface brightness in 
$K_s$ for BCDs and dIs are equally good, indicating that BCD line widths may be enhanced by 
turbulence or winds. 
\end{abstract}


\section{Introduction}

According to \citet{kun00}, the concept of ``compact galaxies'' was introduced by \citet{zwi65}, 
in order to refer to galaxies barely distinguishable from stars on the Palomar Sky Survey plates. 
In a subsequent paper, \citet{zwi70} refined the definition of a ``compact galaxy'' to be any galaxy 
(or part of a galaxy) whose surface brightness is brighter than 20 mag arcsec$^{-2}$ in any chosen 
wavelength range \citep{paz03}. The term ``blue'', as used by \citet{zwi70}, refers to those 
galaxies satisfying the previous condition on both blue and red plates \citep{zwi71}. 
Later on, \citet{thu81} introduced the term ``blue compact dwarf'' (BCD) referring to those 
compact galaxies characterized by three main properties: low luminosity ($M_B \gtrsim -18$), 
small sizes (diameters less than 1 kpc), and strong narrow emission lines superposed on a nearly 
flat continuum, similar to HII regions in spiral galaxies. As their name suggests, the visible 
colors of BCDs appear to be blue. The star forming rates in BCDs are huge in comparison with those 
in dIs (between 0.1 and 1 $M_{\odot}$ yr$^{-1}$, e.g. \citealp{fan88}), and metallicities are low 
(oxygen abundance between $Z_{\odot}/50$ and $Z_{\odot}/2$, e.g. \citealp{kun00}), suggesting young 
chemical ages. 

\citet{thu81} undertook the first systematic approach to study BCDs, observing HI 
fluxes for a large sample (115 targets) for which they derived HI profiles, neutral hydrogen masses 
and total masses. Using the IRTF 3m telescope, \citet{thu83} observed in the near-inrared (NIR) a 
smaller BCD sample (36 targets), deriving a color-metallicity relation and a well defined 
color-luminosity relation which suggested that star formation processes are intimately linked to 
total masses. 

\citet{jam94} pioneered NIR surface photometry, observing 13 dwarfs in Virgo including a few BCDs. 
The author concluded that the NIR colours of the BCDs are paradoxically red, possibly indicating 
an intermediate-age stellar population. No simple evolutionary link between BCDs, dIs and dEs in 
the sample could be discerned. 

Using a 3-component decomposition scheme applied to the surface brightness profiles of 12 BCDs, 
\citet{pap96b} found that the size and luminosity of the purported starburst component depends upon 
the luminosity of the underlying component, as well as on the HI mass of the BCD. Comparing the 
structural properties of BCDs with those of other types of dwarfs, the authors found that for the 
same $B$ luminosity, the underlying component of a BCD has a central brightness brighter by $\sim$ 
1.5 mag and an exponential scale length smaller by a factor of $\sim$ 2 than that of dIs and dEs. 

Employing some larger samples, other authors have derived total magnitudes, surface brightness 
profiles and profile decompositions in the visible \citep{dou97,dou99,cai01} and more recently 
in the NIR \citep{bos97,gav00,cai03,noe03,noe04}. 
Based upon a small sample (12 targets) observed in the NIR, \citet{dou01} found a systematic 
excess of light in the $K$ band with respect to $H$ and $J$ and their previous observations in 
the visible, indicating that a stellar population of red giants dominates the flux. Moreover, 
comparisons of the metallicity-color relations of BCDs and globular clusters show very little 
difference, indicating that BCDs are most probably old, in contradiction with previous results. 
Employing a small sample (9 targets) observed in the NIR, \citet{cai03} found that the morphology 
of BCDs is basically the same as in the optical. 

Using HST and ground-based optical and NIR data, \citet{pap02} investigated I Zw 18, 
known to be the least chemically evolved star-forming BCD galaxy in the local Universe, with 
an oxygen abundance of $12+\log$(O/H)$\sim$7.2, or 1/50 of the Sun \citep{sea72,izo99,kun00}. 
In the optical, its exponential profile is not due to an evolved stellar disc underlying its 
star-formation regions, but rather, due to extended ionized gas emission. The unresolved stellar 
component, though very compact, is not exceptional with respect to structural properties for 
intrinsically faint dwarfs. However, it differs strikingly from the red low surface brightness 
host of standard BCDs. Also, unlike more evolved BCDs, the stellar component of I Zw 18 is much 
more compact than the ionized gas envelope. 

Despite three decades of study at various wavelengths, at least two basic questions remain 
open about BCDs, namely their ages and possible evolutionary links between BCDs, dIs and dEs. 
In particular, how to decouple the diffuse underlying component from the young starburst 
regions remains a fundamental problem in studying dwarf galaxies. The solution is easier to 
achieve in the NIR, which has the advantage of better tracing the diffuse component of BCDs, 
whose flux in the visible is overwhelmed by young starbursts. 

About four years ago we started a long-term research programme to investigate the NIR properties 
of dIs and BCDs \citep{vad05b}. The first paper of this study, hereafter Paper I, addressed the dIs 
\citep{vad05a}. Based upon deep data derived from a sample of 34 field dIs observed at CFHT, 
we have recently found an universal law fitting their NIR surface brightness profiles. We also 
found that these data permit the definition of a fundamental plane for the dIs (Paper I). 
The second paper of this study 
approaches the BCDs. To this end, we have acquired NIR deep images of 16 BCDs in the Virgo cluster. 
Using these data and our insights about dIs, we employ surface photometry to address the NIR 
morphology of BCDs in an attempt to provide new insight regarding possible evolutionary links 
between dIs and BCDs. 

The organization of the present paper follows the one of Paper I. In \S \ref{sample} we present 
the sample selection criteria, and in \S \ref{observations} our observations. The image 
reduction method is presented in \S \ref{reduction}. Surface photometry will be addressed in 
\S \ref{surface}. Photometric data are used in \S \ref{properties} to study the NIR properties 
of BCDs. Conclusions are presented in \S \ref{conclusions}.

\section{The Sample}
\label{sample}

The BCD candidates were selected according to the following criteria:

\begin{enumerate}
\item
common distance, i.e. members of the Virgo Cluster as defined by the Virgo Cluster Catalog 
(VCC; \citealp{bin85,bin87,bin93}); 

\item
classified as BCD or dI/BCD, according to the literature (i.e., NED); 

\item 
availabilty of oxygen abundances, preferably from the [\ion{O}{3}]$\lambda$4363 
emission line; 

\item 
low-density environment, as delineated by the X-ray map provided by the ROSAT 
All-Sky-Survey \citep{lee01}; 

\item
absolute magnitude between $-12 \lesssim M_B \lesssim -16$. 
\end{enumerate}

By restricting targets to the Virgo Cluster, we hoped to reduce scatter in distance-dependent 
parameters. By minimizing the density of intergalactic gas, we would avoid perturbations which 
might be caused by ram-pressure stripping \citep{lee03}. Finally, by restricting the luminosity 
range, we hoped to be able to make a statistically relevant comparison with dIs. 
We identified 16 BCDs which satisfy these conditions. They are listed in Table~\ref{tbl-2}. 

In Paper I, it was shown that NIR surface brightness profiles of dIs flatten toward the centre. 
However, two nearby galaxies classified as dIs, namely NGC~1569 and NGC~3738, display 
centrally-peaked surface brightness profiles. NGC~1569, which has a distance modulus of $DM=26.37$ 
(Paper I), is well known for its central starburst activity (e.g., \citealp{gre98,alo01}). NGC~3738, 
which is located in the Canes Venatici I cloud at $DM=29.27$, also has been previously suggested 
to harbour stochastic star formation processes \citep{hun80}. Due to their NIR flux excesses at 
small radii, we add NGC~1569 and NGC~3738 to the Virgo BCD sample in this paper. We will show 
that their surface brightness profiles suggest that they are BCDs, not dIs.

\section{Observations}
\label{observations}

Between 2001 and 2004 we had 4 observing runs (21 nights in total) on the 2.1~m telescope 
at the Observatorio Astron\'{o}mico Nacional in the Sierra San Pedro M\'{a}rtir (OAN-SPM) 
in Baja California, Mexico. The primary objective of the runs was to obtain deep imaging of a 
sample of Virgo BCDs in the NIR, all of which have known or soon-to-be-known metallicities. 

All galaxies were observed in $J$ and $K_s$ bands using the CAMILA NIR camera at the $f/13$ 
focus of the 2.1~m telescope. CAMILA is equipped with a NICMOS3 $256 \times 256$ pixel array
\citep{cru94}. The scale was $0\farcs85 \, \rm pix^{-1}$, yielding a field of view FOV 
$3\farcm6 \times 3\farcm6$. 

The weather during Feb-Mar at OAN-SPM was unstable throughout the four runs, with clear skies 
less than half the allocated time and only one night with photometric conditions. Most of the 
allocated observing time was bright (Moon within 5 days of full), but this is less important when 
observing in the NIR, as moonlight brightens the sky by about 0.3 mag arcsec$^{-2}$ in $K_s$ and 
about 0.9 mag arcsec$^{-2}$ in $J$\footnote{Cf. Vanzi \& Hainaut (2002), online at European 
Southern Observatory, http://www.eso.org/gen-fac/pubs/astclim/lasilla/l-vanzi-poster}. 

During the four runs, we imaged in $J$ and $K_s$ 16 BCDs in Virgo. We took the data starting 
around local time 23h when the airmass was below $\sim1.8$. 
Because of cirrus clouds most of the time, variable exposure times of 20--70 min in $K_s$ and 
15--40 min in $J$ were used in order to reach $\mu_{K_s}=23$ mag arcsec$^{-2}$ and $\mu_J=24$ mag 
arcsec$^{-2}$. These exposures were obtained by combining sequences of one-minute frames in both 
$K_s$ and $J$. These one minute sub-exposures were acquired unguided. 
During the first run, we cycled the targets (VCC 24 and VCC 324 in $K_s$) through 
the four quadrants of the chip. During the next three runs, we sampled the sky half as often as 
the targets in regions chosen randomly to be 4 to 5 arcmin away from the galaxies. Each galaxy frame 
was dithered by about 3$\arcsec$. In short, we adopted the following observing sequence: 
\begin{eqnarray}
sky-target-target-sky-target-target-sky-...-sky-target-target-sky
\end{eqnarray}
We present the observing log for all four runs in Table~\ref{tbl-2}. In the last column we include 
the full width at half maximum (FWHM), measured on a few stars in each reduced image. 

We observed NGC~1569 and NGC~3738 on the Canada-France-Hawaii Telescope (CFHT) in Mar 2002, 
using exposures tuned to match the same surface brightness limit as our OAN-SPM Virgo observations. 
Details of the observations and data reduction of these two galaxies are given in Paper I.

\section{Data Reductions}
\label{reduction}

One bad pixel map was built for each run using flat field images taken with two different 
exposure times. It was applied as the first step in the data reduction process to all our raw 
images (using the BADPIX task of IRAF). 

For each filter, a flat field was built for each run from two series of twilight sky images 
taken with long and short exposure times. The low-signal flat was subtracted from the high-
signal flat in order to produce the final flat field image. 

Removal of the background represents the most important step in NIR observations, especially 
when observing the outer regions of faint galaxies (e.g., \citealp{vad04}). In order to remove 
the variable airglow and thermal contributions, the sky frame observed close to each science 
frame was subtracted from the science frame. The sky-subtracted galaxy frames were divided by the 
flat field, then leveled additively by forcing the corners to zero. Finally, they were combined  
using the IMCOMBINE task in median mode, which eliminated the hot pixels, cosmic rays, and negative 
residuals that appear in the reduced frames after sky subtraction. 

The collection of IRAF scripts created to reduce our data is available online\footnote{The 
REDNIR.CL NIR image reduction scripts available at http://www.geocities.com/ovidiuv/astrsoft.html}. 
The reduced $J$-band images of the 16 Virgo BCDs observed are presented in the left panel of 
Fig.~\ref{OANSPM_galaxies}. 

Throughout our runs, we planned to observe Persson reference stars \citep{per98} to calibrate 
our observations. As only one night turned out to be photometric, we derived the zero-point for 
each reduced image using 3 to 12 2MASS stars appearing in each field. For most images, we estimate 
the zero-point errors to be less than 0.05 mag, with the maximum reaching about 0.2 mag in three 
sparsely populated cases: VCC 641 (including only one 2MASS star in the field), and VCC 144 and 
VCC 1313 (with only two 2MASS stars each).

\section{Surface Photometry}
\label{surface}

Due to the relatively small telescope aperture and poor spatial resolution, it was not possible 
to resolve stars at the distance of Virgo. We employed the task ELLIPSE of the STSDAS package 
under IRAF to perform surface photometry. $K_s$ observations were more affected than $J$ by the 
poor weather conditions (high humidity and cirrus), so we preferred to use images in $J$ to 
determine the ellipse fitting parameters. 

We used the ELLIPSE task in two stages. First, we approximated the initial ellipse centres, 
ellipticities and position angles, but allowed them to vary freely with radius during the fitting. 
Then we plotted the fits using ISOPALL to analyze the outer isophotes where the underlying diffuse 
component is expected to reflect the geometry of the galaxy. At large radii (e.g., close to the 
semimajor axis given by NED\footnote{NASA/IPAC Extragalactic Database (NED) is operated by the 
JPL, CALTECH, under contract with NASA}), the fitting parameters can be regarded as constant, 
none of them showing isophotal twists or tidal distortions in the outer regions. 
Using these values, in the second stage we fixed the centre, ellipticity, and the position angle 
and repeated the fitting process for both bandpasses. Setting the fitting parameters to be constant 
is a more robust approach to model a galaxy's brightness in its outer regions. 
In Table~\ref{tbl-1}, we include the measured eccentricities and position angles of our targets. 

In Fig.~\ref{OANSPM_galaxies}, the upper graphs in the right panel beside each galaxy present 
surface brightness profiles in $K_s$ and $J$ for the BCDs observed at OAN-SPM. The profiles of 
NGC~1569 and NGC~3738 are included in the last two panels. 
The formal uncertainties in the surface photometry as given by ELLIPSE are plotted as error bars. 
Most of the errors are less than 0.1 mag arcsec$^{-2}$. Ellipticity errors listed by ELLIPSE are 
less than 0.05 in most cases, and position angles are uncertain by about 2 degrees. For the faintest 
or the most compact targets (VCC~428, VCC~1313), we estimate that errors reach 0.1 in ellipticity 
and 5-10 degrees in position angle.

\subsection{Isophotal Magnitudes}

Using the ISOPLOT task of STSDAS, we have measured the {\it isophotal magnitude} ($m_I$) of 
each BCD from the total flux integrated out to the faintest visible isophote, i.e., about 
$\mu_{K_s}=23$ mag arcsec$^{-2}$ or $\mu_J=24$ mag arcsec$^{-2}$. We include the isophotal 
magnitudes in Table~\ref{tbl-4}. The errors in the isophotal magnitudes are determined mainly 
by the zero points, whose typical uncertainties are less than 0.1 mag (see \S \ref{reduction}). 

The 2MASS\footnote{2MASS GATOR Catalog http://irsa.ipac.caltech.edu/applications/Gator} extended 
source catalog includes 8 BCDs in common with our sample. Also, Goldmine\footnote{GOLDMine - 
Galaxy Online Database Milano Network http://goldmine.min.infn.it} \citep{gav03} lists NIR data 
for 9 BCDs observed by us. In Table~\ref{tbl-3} we compare the total magnitudes from these 
sources with our own. In most cases, both 2MASS and GOLDMine fluxes are below our fluxes, 
likely due to our fainter detection. The difference is larger for 2MASS, as much as 1 mag for 
VCC~848 (for which GOLDMine $K_s$ data is much closer to our result).

\subsection{Profile Decomposition}

The exponential law, 
\begin{equation}
I = I_0 \exp (-r/r_0)
\end{equation} 
\noindent
and the de Vaucouleurs law, 
\begin{equation}
I = I_e \exp \left(-7.67 \left((r/r_e)^\frac{1}{4} - 1\right) \right)
\end{equation} 
\noindent
have been used extensively to model the surface brightness profiles (SBPs) of BCDs both in 
the visible and in the NIR. Here, $I$ stands for the brightness at a given semimajor axis $r$, 
$I_0$ for the central surface brightness, $I_e$ for the surface brightness at the effective 
radius, while $r_0$ represents the exponential and $r_e$ the effective radius within which 
half of the galaxy's total luminosity is contained. 

\citet{jam94} fitted an exponential in order to model the luminosity profiles of five Virgo 
BCDs observed in the NIR, deriving exponential scale lengths and extrapolated central surface 
brightnesses. Analysing a sample of 44 BCDs observed in the visible, \citet{dou97} and 
\citet{dou99} found that 25\% of the SBPs fit a pure exponential profile, 18\% have a dominating 
exponential component, 20\% follow a pure de Vaucouleurs profile, 27\% present a dominant de 
Vaucouleurs distribution, and the remaining 7\% are unclassifiable. Using a subsample of 12 BCDs 
observed in the near-infrared, \citet{dou01} conclude that the NIR light distributions are generally 
consistent with those in the optical, although in some peculiar cases the brightness profiles 
differ significantly. 

A combination of two or three different models has been employed by some authors to improve the 
fits to the surface brightness profiles. 
\citet{gav00} used the exponential law, the de Vaucouleurs law, or a combination 
of the two in order to fit the light profiles of dIs and BCDs observed in $H$ and $K_s$ in five 
nearby clusters. Using observations in the visible of 12 BCDs and 2 other bright starburst galaxies, 
\citet{pap96b} decomposed profiles into three components: an exponential at large radii, a plateau 
at intermediate radii, and a Gaussian at small radii, with the last two supposedly describing the 
current starburst superimposed on the older stellar component. 

\citet{cai01} performed deep broadband observations in the visible on a sample of 28 BCDs. Over 70\% 
of the galaxies showed complex profiles that precluded fitting with a single standard law (exponential 
or de Vaucouleurs), with extra structure at high to intermediate intensity levels. Using a subsample 
of 9 BCDs, \citet{cai03} performed NIR surface photometry, finding that the morphology is basically 
the same as in the visible, with the inner regions dominated by the starburst component. 
Fits of Sersic laws were very sensitive to the selected radial interval. Fitting an exponential model 
gave more stable results. 

\citet{noe03} and \citet{noe04} took deep NIR ($J$, $H$ and $K_s$) images of a sample 
of 23 field BCDs with the Calar Alto 3.6m telescope in the North and the ESO 3.6m NTT in the 
South. To perform the surface photometry, the authors employed a complicated method \citep{pap02} 
which uses isophotal masks defining ``polygonal apertures'' which follow the flux. 
This method is sensitive to the sky noise at large radii, to the imperfect manual cleaning of the 
field stars, and also to the starburst region at small radii where most of the SBPs 
are seen to brighten abruptly. Nevertheless, exponential fitting was found to approximate well the 
intensity distribution at large radii. The surface brightness profiles of some BCDs 
show a central flattening, the so-called ``type V'' profile, following the nomenclature of 
\citet{bin91}. To describe this distribution, the authors employed the ``modified exponential 
distribution'' - the $med$ law \citep{pap96a} - a complicated function involving two more constants, 
$b$ and $q$, additional to the central intensity, $I_0$, and the exponential scale length, $\alpha$. 
However, a Sersic law was found to model the flattening at small radii, too, although a pure or 
modified exponential formula may be preferred to fit the underlying low surface brightness component 
of BCDs. 

\begin{figure}
\epsscale{0.7}
\plotone{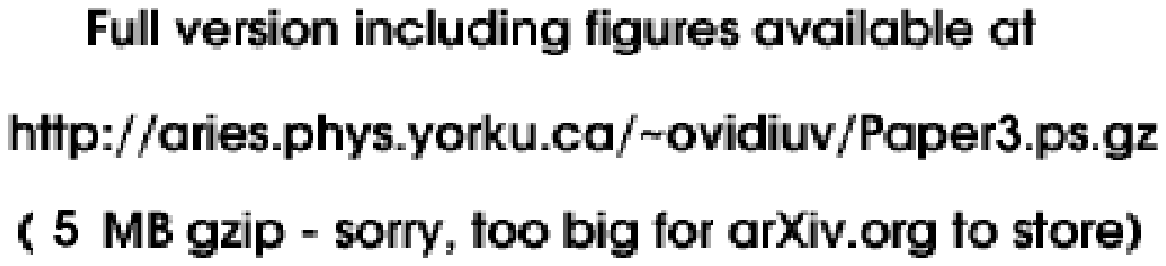}
\label{OANSPM_galaxies} 
\caption{Virgo BCD galaxies observed at OAN-SPM. 
Left panel: $J$ images (North up, East to left, field $3\farcm0\times3\farcm0$). 
Right panel: surface brightness profiles in $J$ and $K_s$ (upper graphs), and 
$J-K_s$ color profiles (lower graphs). To make them more visible, the surface brightness 
profiles in $J$ have been shifted by one magnitude. Solid lines passing through the points 
represent total (sech plus Gaussian) fits. For $K_s$ only, sech and Gaussian components are 
plotted individually as well, with a dashed line and a dotted line, respectively. 
For VCC~428, VCC~1374, VCC~1699 and VCC~1725, the sech component is plotted as a solid line,
because that was the only component fitted. Error bars are those calculated by ISOPLOT. 
NGC~1569 and NGC~3738, star burst dIs with BCD-like profiles observed at CFHT (Paper I) are 
included in the last two panels. 
}
\end{figure}

As mentioned previously, we derived our surface brightness profiles by fixing the ellipse parameters. 
Tying the profiles to the geometry of the smooth extended component instead of the light affected by 
the outburst region is a more robust approach to studying what underlies BCDs. We present our profiles 
in the right panels of Fig.~\ref{OANSPM_galaxies}. Most galaxies show an exponential profile in the 
outer regions, 
extending as low in surface brightness as $\mu_{K_s}$ = 23 mag arcsec$^{-2}$ and $\mu_J$ = 24 mag 
arcsec$^{-2}$. At small radii, superposed on top of the exponential, most profiles show some excess 
over an exponential due to the starburst. Profiles often appear to level off at radii very close to 
zero. These trends have been found previously in dIs, suggesting some structural links with probable 
evolutionary implications for the two types of dwarfs \citep{pap96b}. 

As shown in Paper I, the NIR surface brightness profiles of dIs can be modeled using the following 
hyperbolic secant ($sech$) law: 

\begin{equation}
\label{sech_fit}
I_S = I_{0S} \hbox{ sech} {(r/r_{0S})} = \frac{I_{0S}}{ \cosh (r/r_{0S})} = \frac{2I_{0S}}{e^{r/r_{0S}}+e^{-r/r_{0S}}}
\end{equation}

\noindent
Here $I_S$ represents the fitted $sech$ flux at radius $r$, the distance from the centre along the 
semimajor axis. $I_{0S}$ is the {\it sech central surface brightness} ($m_{0S}$ in magnitude units), 
and $r_{0S}$ represents the {\it sech scale length}. 

We attempted to fit the profiles of BCDs using Equation~\ref{sech_fit} in combination with the 
following Gaussian component to model the starburst: 

\begin{equation}
\label{Gauss_fit}
I_G = I_{0G} \exp {\left(- \frac{1}{2} \left( \frac{r}{r_{0G}} \right)^2 \right)}
\end{equation}

\noindent
Here $I_G$ represents the fitted Gaussian flux at radius $r$, while $I_{0G}$ is the 
{\it Gaussian central surface brightness} ($m_{0G}$ in magnitude units), and $r_{0G}$ represents 
the {\it Gaussian scale length}. The total fitted flux at radius $r$ can be expressed as the sum 
of the two components: 
\begin{equation}
\label{total_fit}
I_T = I_S + I_G
\end{equation}

\noindent
Integrated over all radii, Equation~\ref{sech_fit} defines the {\it sech magnitude}, $m_S$, which 
we interpret as representative of the diffuse underlying component. Toward the centre, the sech 
profile flattens and levels out at the {\it sech central surface brightness}, $\mu_{0S}$. The 
integral of Equation~\ref{Gauss_fit} determines the magnitude of the outburst, $m_G$. 

To perform the fitting of the profiles generated by ELLIPSE, we employed the task NFIT1D (in the 
FITTING package of STSDAS), entering Equation (\ref{total_fit}) as our user specified function 
for the USERPAR/FUNCTION parameter. Fitted surface brightnesses were converted into magnitude 
units (mag arcsec$^{-2}$) using the zero-points for the frames. 

The fits to our surface brightness profiles in $J$ and $K_s$ are shown on the upper graphs in 
the right panels of Fig.~\ref{OANSPM_galaxies}. NGC~1569 and NGC~3738 are included in the last two 
panels. For $K_s$ only, the two components are plotted individually as well, with a dashed line for 
the sech component and with a dotted line for the Gaussian. The fits (solid lines passing through 
the points) match the profiles very accurately from the core to the farthest regions.The sech 
component matches the total at large radii (where the solid and the dashed lines superpose), while 
the Gaussian burst contributes mostly in the cores. Note that the peak surface brightness of the 
Gaussian is as high as that of the underlying sech in most cases. Fits to NGC~1569 and NGC~3738 
are much improved over those with a pure sech function (Paper I). Adopted solutions for parameters 
are summarized in Tables~\ref{tbl-4} and \ref{tbl-5}. 

Due to the poor resolution and seeing, most of the Gaussian scale lengths are comparable with 
the seeing (columns $r_{0G}$ from Tables~\ref{tbl-2} and $FWHM$ from \ref{tbl-4}, respectively). 
In these cases, the Gaussian scale lengths have to be regarded with caution. 
For a few compact galaxies, sech fitting was biased by a central spike in surface brightness, 
heading to an over-estimate of the total flux. In these cases, to prevent an inaccurate fit of 
the exponential at large radii and to account for the seeing (2 to 3\arcsec, cf. with 
Table~\ref{tbl-2}), we restricted the fit to regions beyond about 3\arcsec from the centre. 
Such intervals are marked by horizontal lines in our plots in Fig.~\ref{OANSPM_galaxies}. In 
only one case (VCC~848), we used different inner limits for the two bands; both limits are shown, 
corresponding to $K_s$ (top) and $J$ (bottom). 

There are six galaxies which either were observed in poor weather conditions, have very small 
sizes (and, thus, poor sampling), or have unsure BCD classification. Some of these have profiles 
resembling those of dIs, which can be well fitted using the sech law alone (Paper I). Four BCDs 
give negative Gaussian central surface brightnesses, impossible physically. We re-fitted the 
surface brightness profiles of these six galaxies using a sech component alone. In Table~\ref{tbl-6}, 
we compare the sech model (denoted by $S$ in Column 3) with the two-component model (sech+Gaussian, 
denoted by $S+G$). 

Images in $K_s$ for the following targets were especially affected by the poor weather conditions: 
VCC~428, VCC~1313, VCC~2033, and VCC~848. The first three are also compact, so profiles are poorly 
sampled. As a result of the loss of sensitivity, two consequences are likely. First, we likely have 
less extent to fit the sech component due to the loss of the fainter isophotes. Second, combined 
with our low spatial resolution, sharp features may be blurred somewhat, resulting in Gaussian 
parameters that may be less reliable. 

VCC~428 is a very compact galaxy observed under very poor conditions. A pure sech model matches 
$m_T$ better and its rms errors are slightly smaller than those for the two-component model. Thus, 
we adopt the sech fit. $K_s$ images for VCC~848 also were observed under poor conditions, although 
the rms errors in the two-component fit suggest a better fit than in the sech case. 
VCC~1313 is a very compact BCD, so it is not well sampled. The two-component model provides a 
smaller rms than the sech alone, so we adopt it. However, given the galaxy's compact size and a 
surface brightness profile that is shallower than desirable, it is likely that the sech parameters 
of this composite fit are poorly determined and more uncertain that the formal errors indicate. 

For VCC~1374 and VCC~1725, the fitting derives a negative Gaussian central surface brightness 
($\mu_{0G}$ marked as ``n/a'' in Table~\ref{tbl-6}). According to NED, these two galaxies have 
ambiguous BCD classification, VCC~1374 being listed as ``BCD?, IBm: sp'', VCC~1725 as ``SmIII/BCD''. 
For both galaxies, a pure sech model produces better fits (chi-square and rms values are smaller), 
so we adopt it. 
VCC~1699 has a surface brightness profile resembling that of a dI, but its NED classification 
lists it as ``SBmIII'', although it is listed as a BCD by \citet{paz03} and has been included as a 
dwarf HII galaxy by \citet{vil03}. Its pure sech model provides a better match between the $J$ and 
$K_s$ scale radii, so we prefer to adopt this fit instead of the two-component one. 
In Fig.~\ref{OANSPM_galaxies} we show the fit of the pure sech model for VCC~428, VCC~1374, 
VCC~1699, and VCC~1725, and the fit for the two component model for VCC~848 and VCC~1313. 

The structural parameters for at least two galaxies, NGC~3738 and VCC~848, appear to be unusual. 
The radius of their Gaussian component in both $K_s$ and $J$ is at least a factor of two larger 
than that found for all the others. It is therefore not clear whether their Gaussian components 
represent the same phenomenon as in the other galaxies and may explain their sometimes anomalous 
positions in some of the diagrams that follow. 

Included in Tables~\ref{tbl-4} and \ref{tbl-5} are the sech central surface brightness, $\mu_{0S}$, 
the Gaussian central surface brightness, $\mu_{0G}$, in mag arcsec$^{-2}$, the sech scale length, 
$r_{0S}$, and the Gaussian scale length, $r_{0G}$, in arcsec, for both $J$ and $K_s$. 
Based upon the NFIT1D error analysis, the uncertainties in $\mu_{0S}$ and $\mu_{0G}$ are less 
than $\sim0.1$ mag arcsec$^{-2}$. The sech scale lengths in the two bands agree to within about 
0.5 arcsec (about 5\%) for most galaxies, but the Gaussian scale lengths show larger differences, 
possibly due to variations in image quality. 

\subsection{Sech and Gaussian Magnitudes}

The integrated magnitude of the sech component of a BCD can be calculated from 
Equation~(\ref{sech_fit}) by integrating the surface brightness over radii from zero to 
infinity (Paper I): 
\begin{equation}
\label{int_sech} 
m_S = zp_s - 2.5 \log \left( 11.51 I_{0S} r_{0S}^2 (1-e) \right)
\end{equation}

\noindent
Here $I_{0S}$ is the sech central surface brightness, $r_{0S}$ is the sech scale length, $e$ 
represents the galaxy ellipticity, and $zp_s$ is the frame zero-point. 

The flux due to the starburst can be quantified similarly by integrating Equation~(\ref{Gauss_fit}). 
Let $a$ and $b$ be the semimajor and semiminor axis of the isophotes, and $I_G(a)$ the average 
surface brightness of the Gaussian component in an elliptical annulus with area $dA$: 
\begin{equation}
dA = d(\pi a b) = d(\pi a^2 (1-e)) = 2 \pi a (1-e) da
\end{equation} 

\noindent
Then, the total flux $I_G$ due to the Gaussian component is given by: 
\begin{equation}
\label{int_flux} 
I_G = \int_{0}^{\infty}{I_G(a) dA} = 2 \pi (1-e) I_{0G} \int_{0}^{\infty}{ a \exp \left( -0.5 ( a/r_{0G}) ^2 \right) da} 
\end{equation} 

\noindent
By substituting $x=a/r_{0G}$ in the last integral, the integrated Gaussian flux becomes: 
\begin{equation}
\label{int_flux} 
I_G = 2 \pi (1-e) I_{0G} r_{0G}^2  \int_{0}^{\infty}{x \exp \left( \frac{-x^2}{2} \right) } dx = 2 \pi (1-e) I_{0G} r_{0G}^2 
\end{equation}

\noindent
With this result, the {\it Gaussian magnitude} can be calculated from: 
\begin{equation}
\label{int_mag_Gauss} 
m_G = zp_s - 2.5 \log I_G
\end{equation}

\begin{figure}
\epsscale{0.8}
\plotone{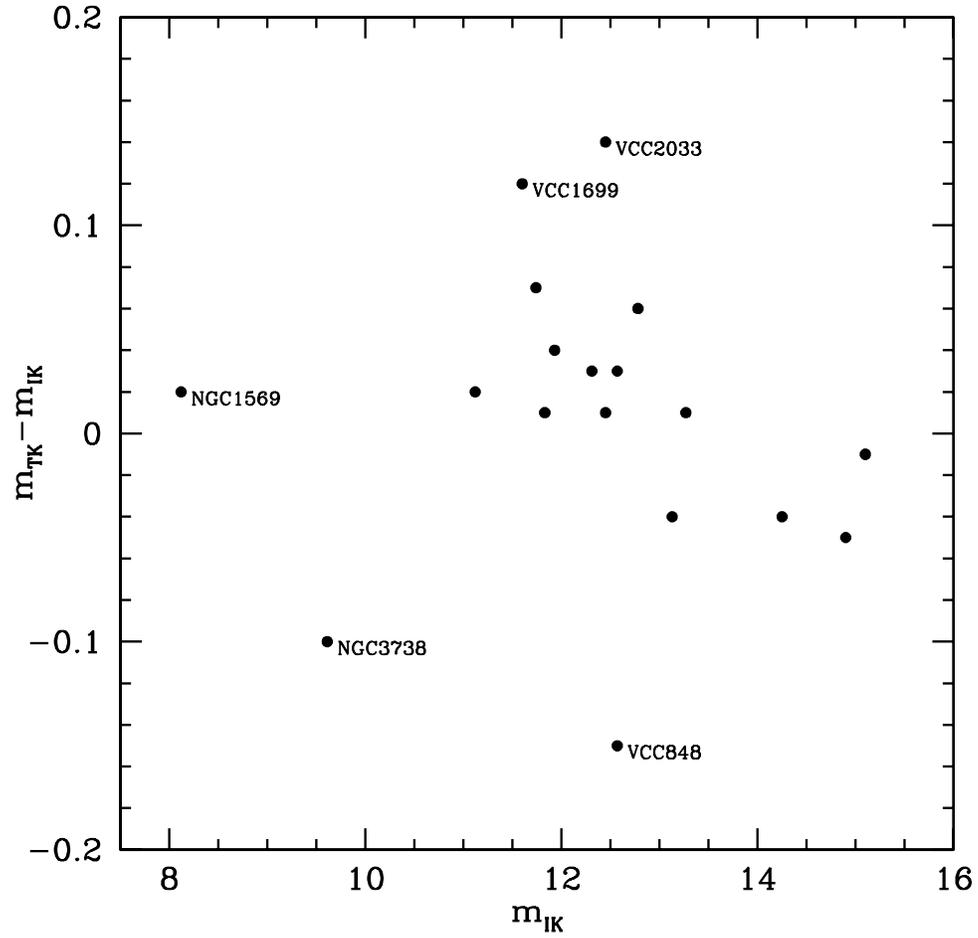}
\caption{
\label{delmag} The difference between the apparent total magnitude $m_T$ and the apparent 
isophotal magnitude $m_I$ versus $m_I$ for the $K_s$ band. The three Virgo BCDs with residuals 
larger than 0.1 mag were all observed in poor weather. 
}
\end{figure}

Included in Table~\ref{tbl-4} for each galaxy are the {\it sech magnitude} ($m_S$), the 
{\it Gaussian magnitude} ($m_G$), and the {\it total magnitude} ($m_T$) computed from the sum 
of the sech and the Gaussian fluxes. Formal uncertainties in $m_S$ and $m_G$ are typically $\sim0.1$ 
mag. As one can see, total magnitudes are very close to the isophotal magnitudes (in most cases 
within 0.05 mag). This is visualized in Fig.~\ref{delmag}, which plots the difference between the 
total magnitude and the isophotal magnitude versus the isophotal magnitude in $K_s$. A few outliers 
are labelled, all of which were observed under particularly difficult circumstances. 

\begin{figure}
\epsscale{0.8}
\plotone{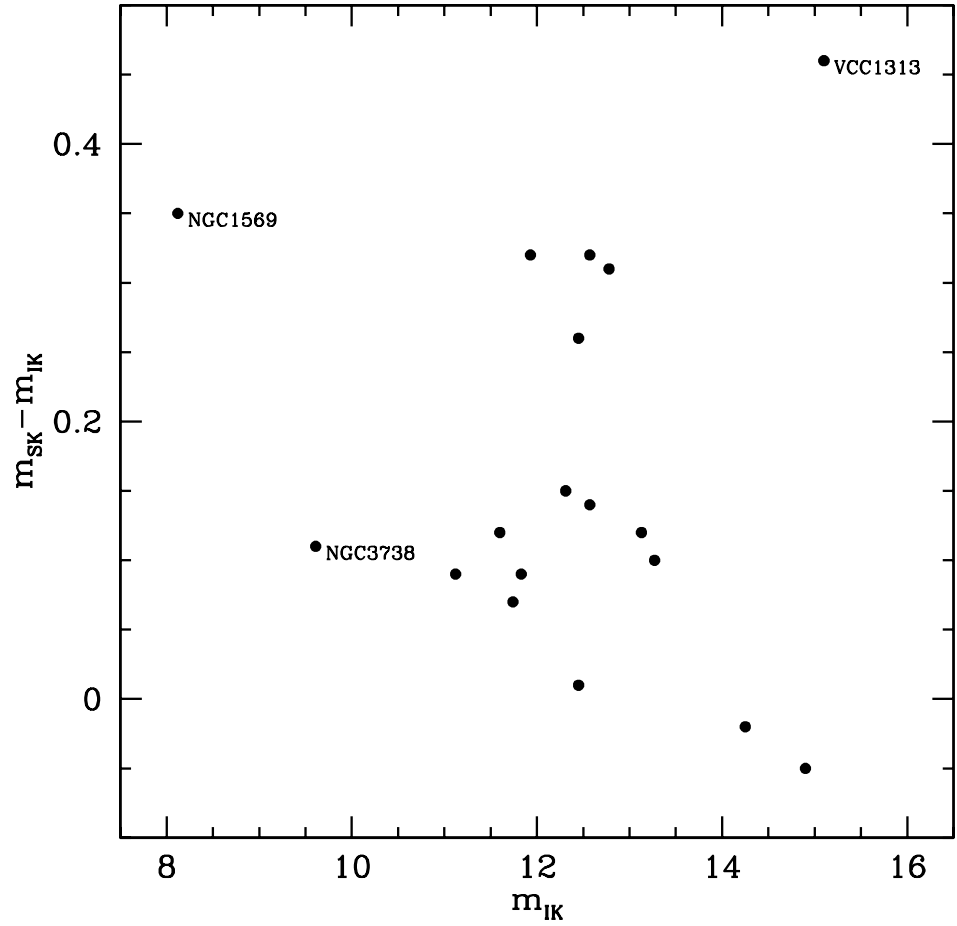}
\caption{
\label{delsmag} The difference between the apparent sech magnitude $m_S$ and the apparent 
isophotal magnitude $m_I$ versus $m_I$ for the $K_s$ band. The graph reveals the effect of 
starbursts on $K_s$ magnitudes. VCC1313 has a very small size, which limited the number of 
samples for fitting. 
}
\end{figure}

In Fig.~\ref{delsmag} we plot the difference between $m_S$ and $m_I$ as a function of isophotal 
magnitude. In $K_s$, sech magnitudes are within about 0.4 mag of the isophotal magnitudes, showing 
that the flux from the starburst is small relative to the total NIR flux. \citealp{sud95} and 
\citealp{sal99} estimated that the starburst in the visible enhances the flux by $0.6-1.0$ mag 
and $0.75$ mag, respectively, on average. Thus, the relative visibility of the underlying component 
is improved in the NIR. 

\subsection{Isophotal Radii}

Using Equation (\ref{sech_fit}), we have calculated the semimajor axis $r_{22}$ corresponding 
to $K_s=22$ mag arcsec$^{-2}$ as well as $r_{23}$ corresponding to $J=23$ mag arcsec$^{-2}$. The 
positive solution for the 2nd order equation in $r_{22}$ comes to 
\begin{equation}
\label{int_rad} 
r_{22} = r_{0S} \hbox{ln} \frac{1+\sqrt{1-A_{22}^2}}{A_{22}} 
\end{equation}

\noindent
where $A_{22}$ is given by: 
\begin{equation}
\log A_{22} = \frac{zp_s - 22 - 2.5 \log I_{0S}}{2.5}
\end{equation}

\noindent
There is a similar formula for $r_{23}$. Our values for $r_{22}$ and $r_{23}$ are listed in 
Table \ref{tbl-4}. Based upon the NFIT1D error analysis, we evaluate typical uncertainties in $r_0$, 
$r_{22}$ and $r_{23}$ to be about 2\%.  

\subsection{Color Profiles}

Color profiles were derived by subtracting $J$ and $K_s$ surface brightnesses via the TCALC task 
of TABLES (STSDAS package). For each galaxy, the color profile is given in Fig.~\ref{OANSPM_galaxies} 
below the surface brightness profile (to the right of the galaxy image). Most galaxies show a 
remarkably constant $J-K_s = 0.7$ to $0.9$ mag. Bursts appear to have little effect on NIR colors.

\section{NIR Properties of BCDs}
\label{properties}

\subsection{Correlations}

Below, we examine correlations among the different parameters of BCDs included in Table~\ref{tbl-4}. 
For sizes, we employ the semimajor axes, $r_{22}$ for $K_s$, and $r_{23}$ for $J$. To trace the diffuse
component, we use the absolute sech magnitudes $M_S$, while for the total light we use the absolute 
isophotal and total magnitudes $M_I$ and $M_T$, respectively. We trace the strength of the starburst 
with the absolute Gaussian magnitude, $M_G$. To characterize the structure of the 
diffuse cores, we employ the sech central surface brightness $\mu_{0S}$ and the sech scale length 
$r_{0S}$, expressed in kpc. We focus on the $K_s$ band because of its reduced sensitivity to the 
light of hot stars, but similar results are obtained in $J$. To compare properties of BCDs with dIs, 
we overlay on some plots our field dI data from Paper I. In each graph to be discussed, typical 
uncertainties in each parameter are plotted as an error cross. For the Virgo galaxies, absolute 
magnitudes and scale lengths are calculated assuming a common distance modulus, DM=30.62 
\citep{fmg01}, which is the HST Key Project distance to the Virgo Cluster anchored to the maser 
distance for NGC 4258 (Paper I). 

\subsubsection{Size and Brightness}

In the visible ($B$ band), emission-line galaxies (mostly BCDs) and gas rich dwarfs display 
a linear relationship between the log of the disk scale length and the absolute magnitude 
\citep{ven00,ber99}. A similar trend has been found in the NIR for dwarf galaxies, including BCDs 
\citep{pap96b,pap02}. The correlation for the BCDs appears to be steeper than the correlation for 
other dwarfs, in the sense that, for a given $B$ luminosity, BCDs have smaller scale lengths by 
a factor of $\sim2$ than other dwarf galaxies \citep{pap96b}. A similar trend was found by 
\citet{bos97} between the NIR concentration index $c_{31}$ (defined as the ratio between the 
radii that contain 75\% and 25\% of light) and the magnitude (see their Fig.~10) for late-type 
galaxies in Virgo. 

\begin{figure}
\epsscale{0.8}
\plotone{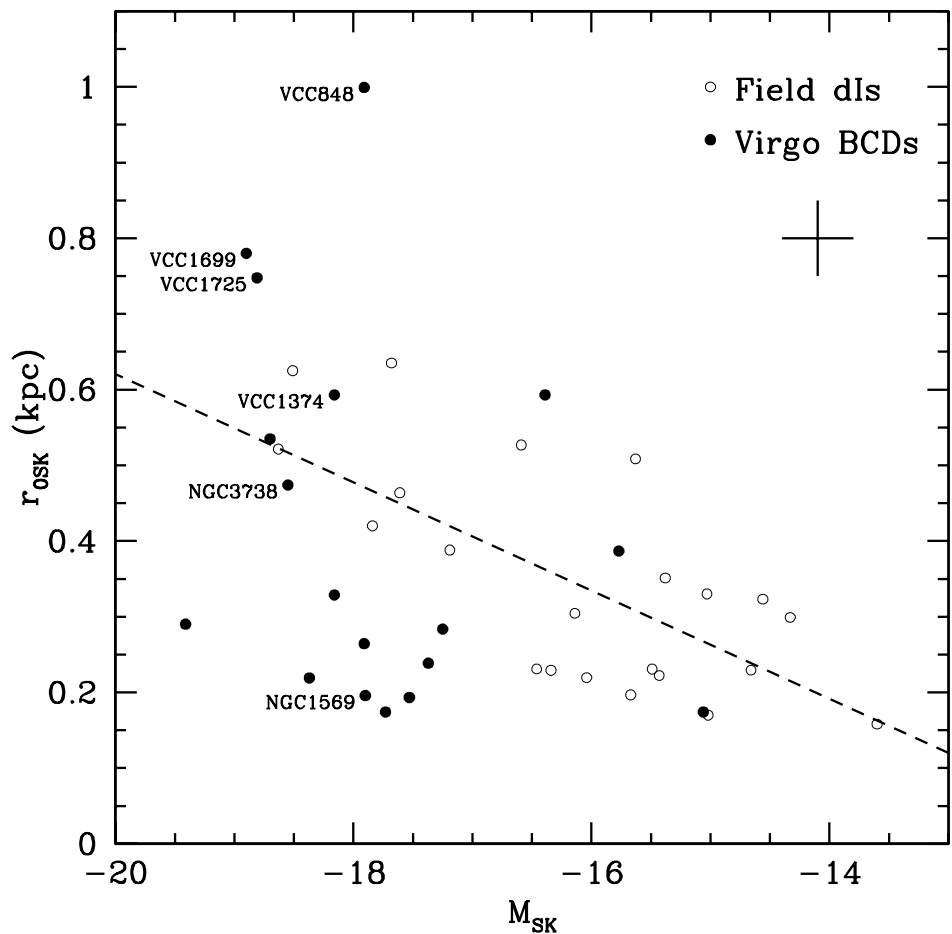}
\caption{
\label{rzermag} The scale length $r_{0S}$ of the sech profile versus the absolute magnitude 
$M_S$ of the sech component in $K_s$. BCDs and the two star burst dIs are represented as filled 
circles, and dIs from Paper I as open circles. The dashed line is a linear fit to the dIs alone. 
VCC 848 appears as an outlier, while VCC~1699, VCC~1725 and VCC~1374 were fitted using the sech 
component alone. 
}
\end{figure}

In Fig.~\ref{rzermag}, we plot the sech scale length $r_{0S}$ versus the sech absolute magnitude 
$M_S$ for our BCD sample (filled symbols). dIs from Paper I are shown as open symbols, and the linear 
fit to the dIs from Paper I is shown as a dotted line. The star burst dIs NGC~1569 and NGC~3738 are 
labeled. VCC~848 is an outlier, whose data was acquired under poor weather conditions. VCC~1699, 
VCC~1725 and VCC~1374 were fitted using the sech component alone. For the brighter BCDs, scale 
length does appear systematically smaller compared to dIs at a given absolute magnitude. 

\begin{figure}
\epsscale{0.8}
\plotone{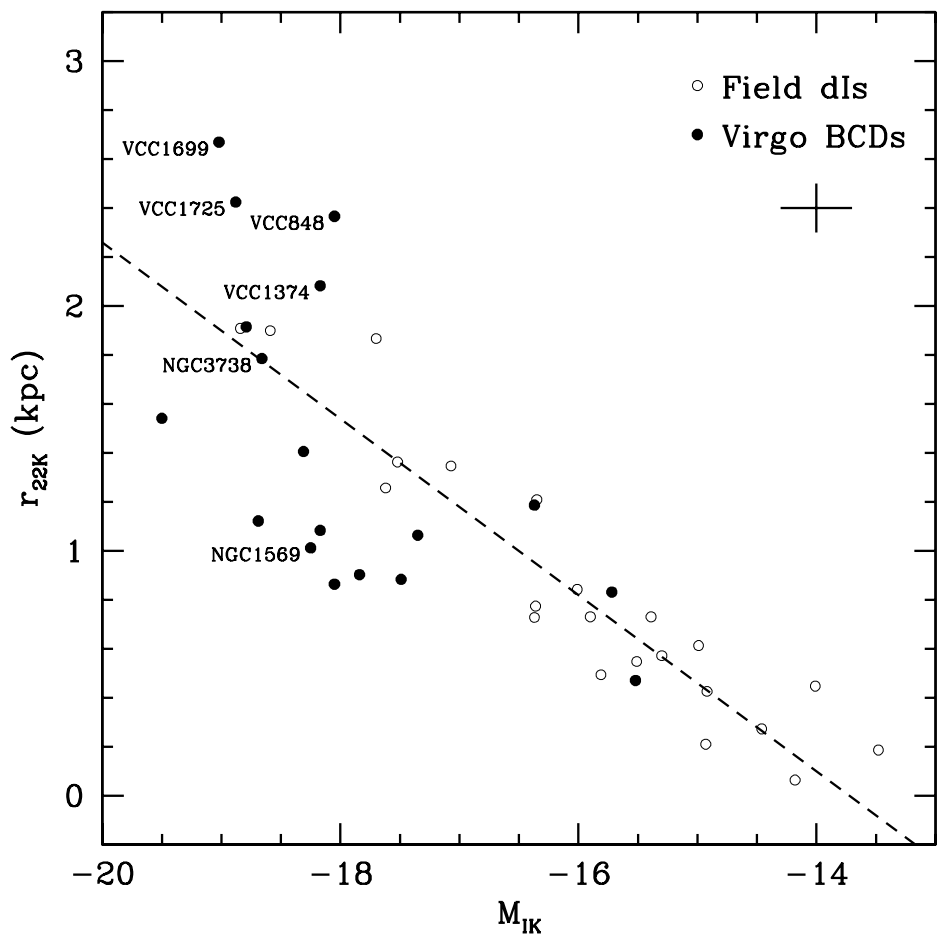}
\caption{
\label{r22mag} The semimajor axis $r_{22}$ of the isophote $K_S$=22 mag arcsec$^{-1}$ 
of the sech component, versus the absolute isophotal magnitude $M_I$ in $K_s$. BCDs and the 
two star burst dIs are represented as filled circles, while dIs from Paper I as open circles. 
Two two starburst field galaxies, NGC~1569 and NGC~3738, are labeled. VCC~1699, VCC~1725, 
and VCC~1374 were all fitted with sech component alone. VCC~848 is an outlier observed in 
poor conditions. A linear fit to the dIs alone is shown as a dashed line. 
}
\end{figure}

Using two BCD samples observed in the visible (22 and 23 targets) and a subsample observed in 
the NIR (11 targets), \citet{dou97} and \citet{dou99,dou01} found a linear trend between the log of 
the effective radius and the absolute magnitude, in the sense that brighter galaxies have larger 
sizes. A similar relation was found by \citet{ven00} in the visible for a larger sample which included 
about 115 BCDs. 
Fig.~\ref{r22mag} plots the semimajor axis $r_{22}$ versus the absolute isophotal magnitude $M_{I}$ 
in $K_s$. BCDs are marked with filled symbols and dIs with open circles. The fit to the dIs derived 
in Paper I is shown as a dashed line. As in Fig.~\ref{rzermag}, VCC~848 and VCC~1699 are outliers. 
The BCDs display a trend similar to that of dIs, albeit with more scatter. The enhanced scatter is 
greater than can be explained by the inclussion of the starburst in $M_I$. 

\subsubsection{Central Surface Brightness}

\citet{ven00} did not find any correlation between the extrapolated central surface brightness 
of an exponential and the absolute $B$ magnitude for a large sample of emission-line galaxies 
(mostly BCDs) observed in the visible. However, using another large sample of dwarf galaxies including 
BCDs, dIs and dEs, \citet{pap96b} and \citet{pap02} found a trend between the central surface 
brightness of the exponential and the total $B$ magnitude, although the correlation was very loose. 

\begin{figure}
\epsscale{0.8}
\plotone{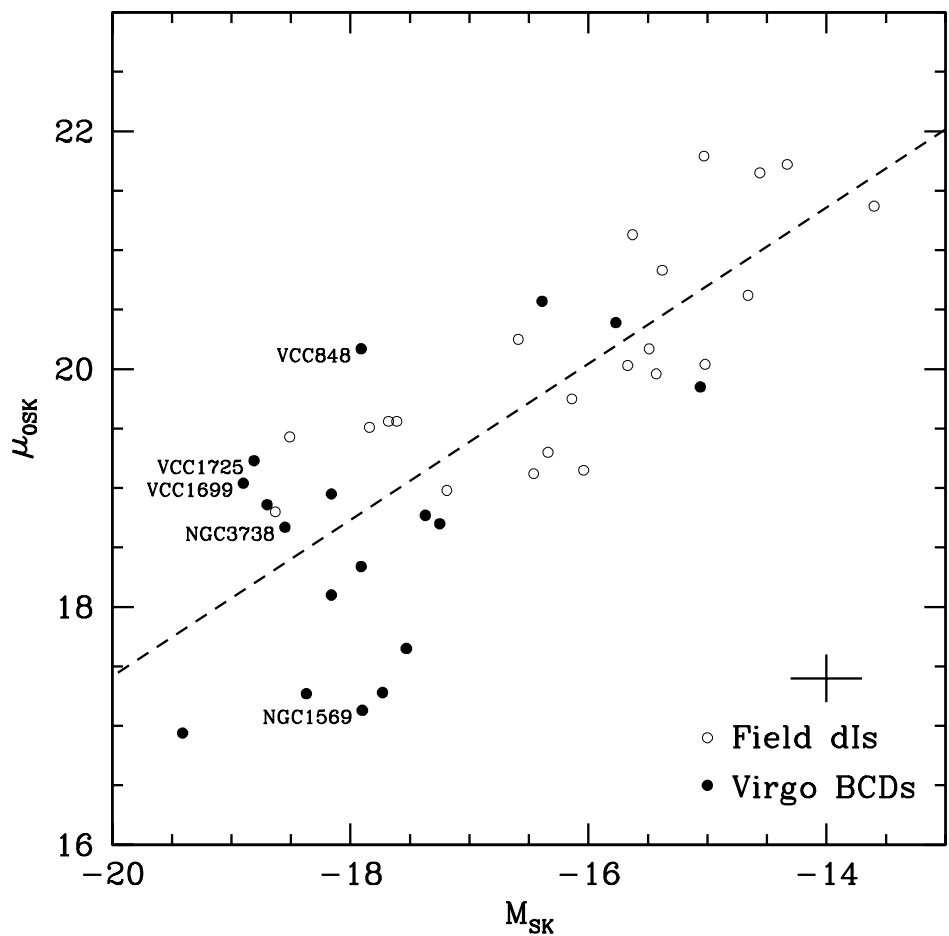}
\caption{
\label{magzmag} The sech central surface brightness $\mu_{0S}$ versus the sech absolute 
magnitude $M_{S}$ for the $K_s$ band. BCDs and the two star burst dIs are plotted with 
filled circles, while dIs from Paper I with open circles. The two star burst dIs are labeled. 
VCC~848 is an outlier observed in poor weather, while VCC~1725 and VCC~1699 were fitted using 
sech component alone. A fit through the rest of the points (dIs and BCDs) is plotted as a 
dashed line. 
}
\end{figure}

In Fig.~\ref{magzmag}, we present the correlation between the sech central surface brightness 
$\mu_{0S}$ and the absolute sech magnitude $M_{S,K}$ in $K_s$. BCDs and the two starburst dIs are 
plotted with filled symbols and dIs from Paper I are plotted with open symbols. BCDs and dIs follow 
similar trends, although luminous BCDs might have systematically brighter central surface brightnesses 
than luminous dIs. The two star burst field dIs are labeled. VCC~848 is an outlier observed in poor 
weather, while VCC~1725 and VCC~1699 were fitted using sech component alone. Using 38 dIs and BCDs 
in our two samples, the following relation is derived: 
\begin{equation}
\label{magzmagfit} 
M_{S,K} = ( 1.52 \pm 0.08) \mu_{0S,K} - ( 46.51 \pm 1.35 )
\end{equation}
\noindent
The fit is shown as a dotted line in Fig.~\ref{magzmag}. The rms deviation of points about the 
fit is 0.84 mag. 

\subsubsection{Starbursts}

\begin{figure}
\epsscale{0.8}
\plotone{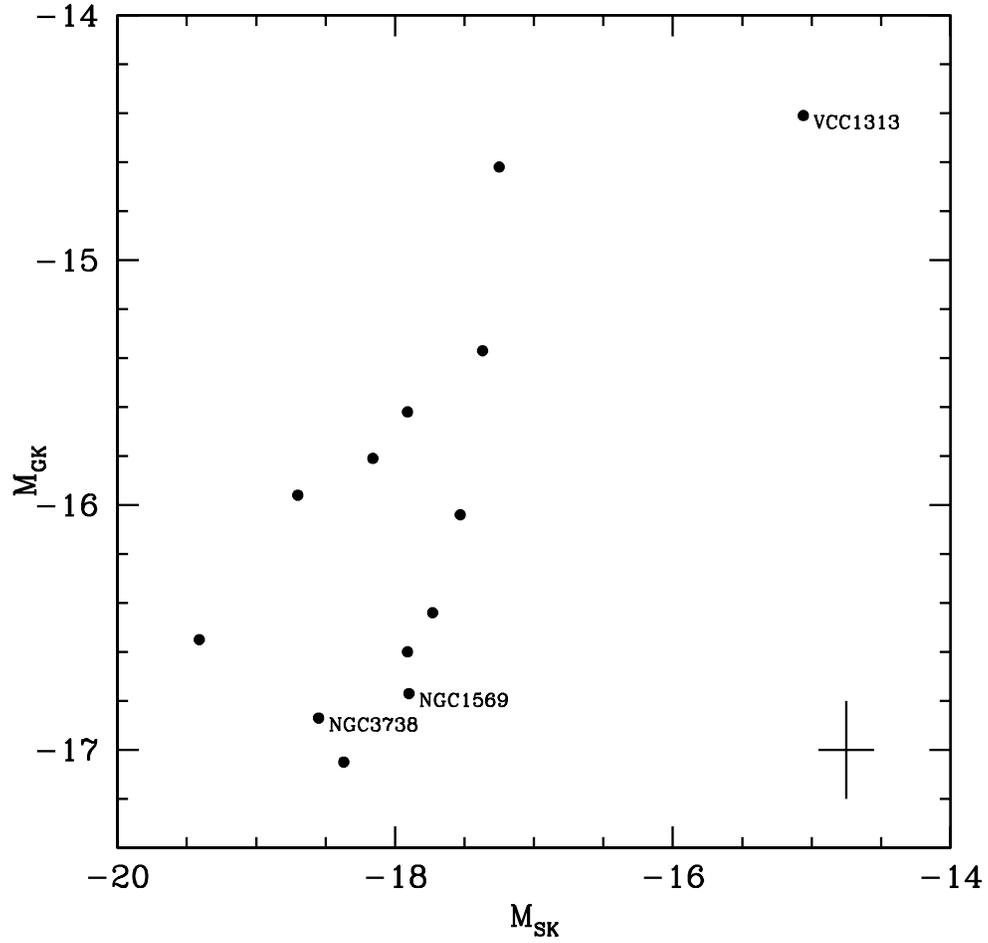}
\caption{
\label{maggmagsa} The Gaussian magnitude $M_{G}$ versus the sech magnitude $M_{S}$ 
for our BCD sample and two star burst dIs, in $K_s$. There is a rough correlation between 
the strength of the starburst and the luminosity of the diffuse component, in the sense 
that objects with brighter underlying components have more luminous bursts. VCC 1313's 
position may well be the result of the poorer quality of the data available for it. 
}
\end{figure}

\begin{figure}
\epsscale{0.8}
\plotone{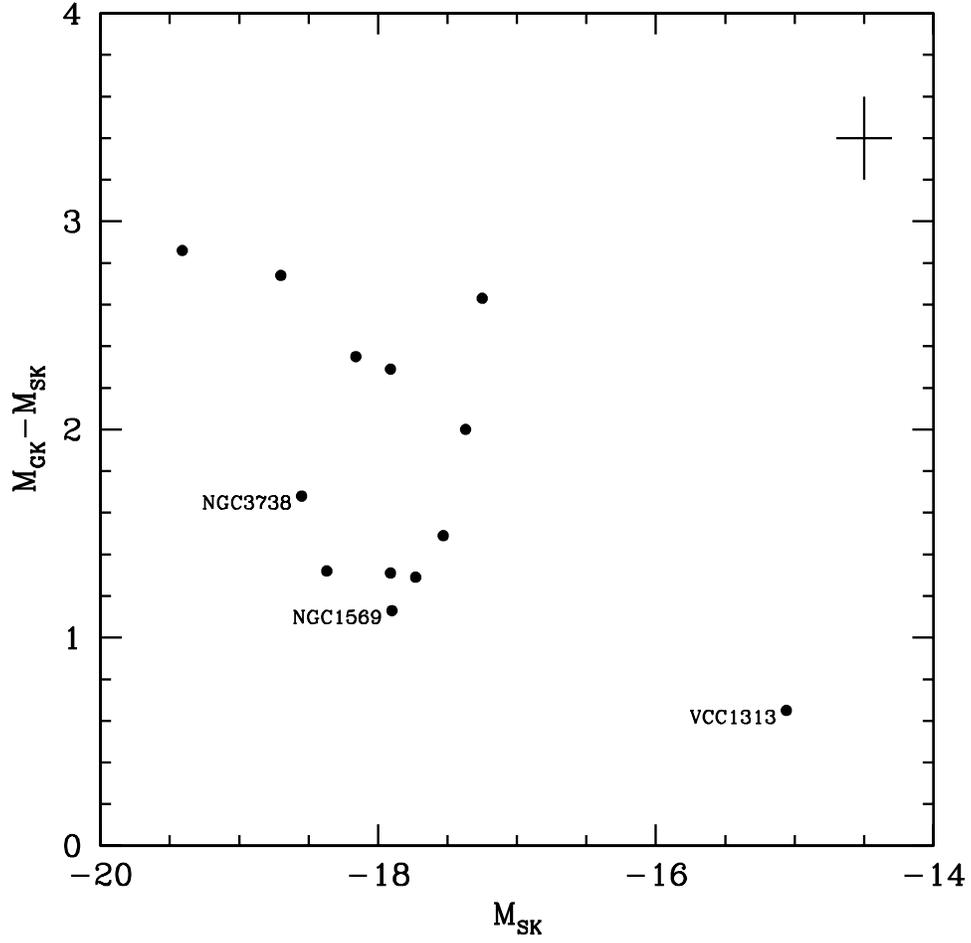}
\caption{
\label{maggmagsb} $M_G-M_S$ versus $M_S$. There is no evidence that the relative strength 
of bursts grows with luminosity. VCC~1313 was observed in poor weather, being also a very 
compact object. 
}
\end{figure}

Fig.~\ref{maggmagsa} plots the Gaussian magnitude $M_{G}$ versus the sech magnitude $M_{S}$ 
for our BCD sample. Objects with brighter underlying components seem to have larger bursts. 
VCC 1313's position may well be the result of the poorer quality of the data available for it. 
Fig.~\ref{maggmagsb} shows the strength of the burst relative to the diffuse component 
($M_{G}-M_{S}$) versus $M_{S}$. There is no obvious corellation, which indicates that burst 
strenght scales linearly with mass. In other words, there is no evidence for triggering of 
star formation in more massive hosts over and above what is expected for a large reservoir 
of matter. VCC 1313's position may well be the result of the poorer quality of the data 
available to it. 

\subsubsection{Fundamental Plane}

\begin{figure}
\epsscale{0.8}
\plotone{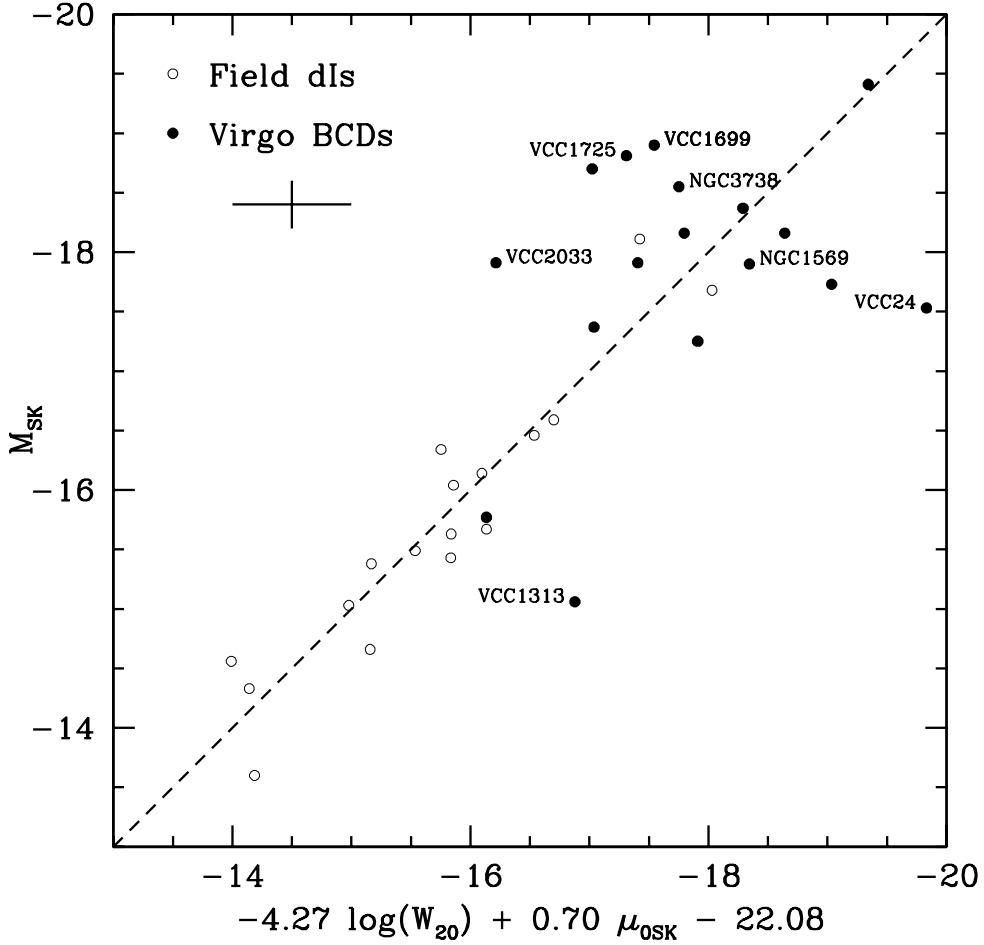}
\caption{
\label{WMZMI} The dI fundamental plane in $K_S$, as described by the relation between 
the absolute sech magnitude $M_S$, HI line-width $W_{20}$, and the central sech surface 
brightness $\mu_{0}$ (Paper I). dIs are represented by open circles, while BCDs and the two 
star burst dIs are marked by filled circles. The BCDs show more spread than the dIs, but 
they scatter around the same relation. The error bar in the X axis reffers to average BCD 
data and is due mainly to uncertainties in $W_{20}$. NGC~1569 and NGC~3738 are labeled, 
along with the most deviant Virgo BCDs, which have the largest uncertainties in $W_{20}$ 
(as high as 13 km/s, which trtanslates into 4.5 mag in the first term of the FP). 
}
\end{figure}

For spiral galaxies, absolute magnitudes are tightly correlated with the log of the HI 
line-width $W_{20}$ after correcting for projection \citep{tul77}. However, the correlation 
is poor for late-type galaxies, even when NIR data is employed (\citealp{pie99}, Paper I). 
In Paper I it was discovered that residuals in the Tully-Fisher relation for dIs were linked 
to the central surface brightness of the sech profile, $\mu_{0S}$. A ``fundamental plane'' 
was defined for the dIs, relating $M_{S}$ to $\mu_{0S}$ and $\log W_{20}$. In Fig.~\ref{WMZMI}, 
we reproduce this fundamental plane, representing dIs from Paper I as empty symbols and 
the BCDs as filled symbols. Although BCDs show more scatter than the dIs, they are concentrated 
around the plane, suggesting that BCDs are dynamically similar to dIs. The two star burst 
field dIs, NGC~1569 and NGC~3738, are labeled, along with the most outlier Virgo BCDs. The 
larger scatter of BCDs about the plane is mainly due to the large uncertainties in $W_{20}$, 
as high as 13 km/s, which translates into 4.5 mag in the first term of the X axis. Also it 
could be due to intrinsic factors affecting the line widths, such as turbulence or winds and 
to differences in distances due to the cluster depth. 

\section{Conclusions}
\label{conclusions}

Blue compact dwarfs (BCDs) and dwarf irregulars (dIs) are important probes for studying the 
formation and evolution of galaxies. In an effort to extract information about old stellar 
populations, a sample of 16 BCDs in the Virgo Cluster have been imaged using a NIR array at 
the OAN-SPM in Mexico. 

Surface brightness profiles were successfully modeled using only two functions (four free 
parameters): a hyperbolic secant (sech) tracing the diffuse component (responsible for most 
of the light), and a Gaussian for the central starburst. A hyperbolic secant is known to fit 
the near-IR profiles of dIs. Isophotal, sech, and total (sech plus Gaussian) NIR magnitudes 
were calculated for all galaxies. Also, semimajor axes at $\mu_J=23$ mag arcsec$^{-2}$ and 
$\mu_{K_s}=22$ mag arcsec$^{-2}$ were determined from the sech fit. 

We searched for relations between semimajor axes, scale length, absolute magnitude (sech, 
isophotal, total, and Gaussian), central surface brightness, and color. Correlations between 
the sech scale length and the sech magnitude, sech surface brightness and sech magnitude, and 
between the sech semimajor axis and the total magnitude overlap those of dIs, though with more 
scatter. Overall, BCDs with more luminous diffuse components are larger and have brighter 
cores, as is also found for dIs. The central burst appears to correlate with the luminosity 
of the diffuse component, with brighter BCDs having stronger star bursts. However, the strength 
of the burst relative to the diffuse component shows no trend with luminosity, indicating that 
more luminous bursts are simply a consequence of a larger reservoir of matter, not extra 
triggering. Color profiles show a relatively constant $J-K_s=0.7$ to $0.9$ mag at all radii. 
The diffuse component represents the overwhelming majority of the NIR light for most BCDs, 
the starburst enhancing the flux by less than about 0.3 mag. 

BCDs lie on the fundamental plane of dIs (Paper I), showing the same relation between the 
sech absolute magnitude, the sech central surface brightness, and the hydrogen line-width. 
However, the scatter about this plane is larger than for the dIs, perhaps due to enhanced 
turbulence or winds. It is concluded that BCDs and dIs are similar structurally and dynamically.

\acknowledgments

We thank the OAN-SPM time allocation committees for granting us the opportunity to observe. 
MGR acknowledges financial support from CONACyT grant 37214-E and DGAPA-UNAM grant IN114199. 
Special thanks to F. Montalvo, 
S. Monrroy, and G. Melgoza, for their help with the observations. MLM thanks the Natural 
Sciences and Engineering Council of Canada for its continuing support. For our data reductions, 
we used IRAF, distributed by the National Optical Astronomy Observatories, which are operated 
by the Association of Universities for Research in Astronomy, Inc., under cooperative agreement 
with the National Science Foundation. This research has made use of the GOLDMine Database in
Milano and the NASA/IPAC Extragalactic Database (NED) which is operated by the JPL, CALTECH, 
under contract with NASA.

\clearpage
\begin{deluxetable}{lcccr}
\tablewidth{0pt}
\tabletypesize{\scriptsize}
\tablecaption{Galaxy Sample \label{tbl-1}}
\tablehead{
\colhead{Galaxy} &
\colhead{$\alpha$ (J2000)} &
\colhead{$\delta$ (J2000)} &
\colhead{$e$} &
\colhead{$PA$} \\
\colhead{(1)} &
\colhead{(2)} &
\colhead{(3)} &
\colhead{(4)} &
\colhead{(5)} 
}
\startdata
VCC 24   & 12:10:35.6 & +11:45:39 & 0.35 & $-$25 \\
VCC 144  & 12:15:18.3 & +05:45:39 & 0.30 & $-$65 \\
VCC 213  & 12:16:56.0 & +13:37:31 & 0.15 & $-$70 \\
VCC 324  & 12:19:09.9 & +03:51:21 & 0.25 & $+$30 \\
VCC 334  & 12:19:14.2 & +13:52:56 & 0.00 &     0 \\
VCC 428  & 12:20:40.2 & +13:53:20 & 0.60 & $+$40 \\
VCC 459  & 12:21:11.3 & +17:38:19 & 0.40 & $+$60 \\
VCC 641  & 12:23:28.4 & +05:48:59 & 0.40 & $-$60 \\
VCC 802  & 12:25:28.7 & +13:29:50 & 0.65 & $+$57 \\
VCC 848  & 12:25:52.5 & +05:48:36 & 0.65 & $+$35 \\
VCC 1313 & 12:30:48.5 & +12:02:42 & 0.40 & $+$73 \\
VCC 1374 & 12:31:38.0 & +14:51:24 & 0.60 & $-$18 \\
VCC 1437 & 12:32:33.5 & +09:10:25 & 0.26 & $+$75 \\
VCC 1699 & 12:37:03.0 & +06:55:36 & 0.50 & $+$55 \\
VCC 1725 & 12:37:41.2 & +08:33:33 & 0.40 & $-$65 \\
VCC 2033 & 12:46:04.4 & +08:28:35 & 0.10 & $+$40 \\
\enddata
\tablecomments{
Col.~(1): Galaxy name; 
Col.~(2): Right ascension (listed by NED); 
Col.~(3): Declination (listed by NED); 
Col.~(4): Ellipticity (1-b/a), from this paper; 
Col.~(5): Position angle of major axis (North through East), from this paper; 
}
\end{deluxetable}

\clearpage
\begin{deluxetable}{ccccr}
\tabletypesize{\scriptsize}
\tablewidth{0pt}
\tablecaption{Observing Log \label{tbl-2}}
\tablehead{
\colhead{Galaxy} & 
\colhead{Date (UT)} &
\colhead{Filter} & 
\colhead{Exp Time} &
\colhead{FWHM} \\ 
\colhead{} &
\colhead{} &
\colhead{} &
\colhead{(sec)} & 
\colhead{(arcsec)} 
}
\startdata
VCC 24   & 2002 Feb 26       & $J$   &  600 & 2.1 \\
         & 2001 Mar 1        & $K_s$ & 3720 & 2.5 \\ 
VCC 144  & 2003 Mar 14       & $J$   & 1200 & 2.0 \\
         & 2003 Mar 19       & $K_s$ & 2340 & 2.2 \\
VCC 213  & 2003 Mar 19       & $J$   & 1200 & 2.3 \\
         & 2003 Mar 19       & $K_s$ & 2160 & 2.5 \\
VCC 324  & 2002 Feb 26       & $J$   &  600 & 2.0 \\
         & 2001 Mar 2        & $K_s$ & 3660 & 2.5 \\
VCC 334  & 2003 Mar 19       & $J$   &  660 & 2.2 \\
         & 2003 Mar 19       & $K_s$ & 1320 & 2.8 \\
VCC 428  & 2004 Feb 17       & $J$   & 1200 & 2.3 \\
         & 2004 Feb 17       & $K_s$ & 2340 & 2.0 \\
VCC 459  & 2004 Feb 12       & $J$   & 1200 & 1.8 \\
         & 2004 Feb 12       & $K_s$ & 2400 & 2.4 \\
VCC 641  & 2002 Feb 26       & $J$   &  600 & 2.2 \\
         & 2003 Mar 17       & $K_s$ & 4500 & 2.6 \\
VCC 802  & 2004 Feb 14,15    & $J$   & 1800 & 2.6 \\
         & 2004 Feb 14       & $K_s$ & 2280 & 2.3 \\
VCC 848  & 2004 Feb 12       & $J$   & 2400 & 2.4 \\
         & 2004 Feb 12       & $K_s$ & 2340 & 2.3 \\
VCC 1313 & 2004 Feb 15,16    & $J$   & 2400 & 2.5 \\
         & 2004 Feb 15,16,17 & $K_s$ & 4560 & 3.0 \\
VCC 1374 & 2004 Feb 13       & $J$   & 1200 & 2.5 \\
         & 2004 Feb 13       & $K_s$ & 2160 & 2.5 \\
VCC 1437 & 2003 Mar 14       & $J$   & 1680 & 1.9 \\
         & 2002 Feb 28       & $K_s$ & 2700 & 2.4 \\
VCC 1699 & 2004 Feb 14,17    & $J$   & 2400 & 2.6 \\
         & 2004 Feb 14,16    & $K_s$ & 4440 & 2.4 \\
VCC 1725 & 2004 Feb 13       & $J$   & 1200 & 2.6 \\
         & 2004 Feb 13       & $K_s$ & 2340 & 2.6 \\
VCC 2033 & 2003 Mar 14       & $J$   & 1200 & 2.0 \\
         & 2002 Feb 28       & $K_s$ & 1200 & 1.8 \\
\enddata
\end{deluxetable}

\clearpage
\begin{deluxetable}{lcccccc}
\tabletypesize{\scriptsize}
\tablewidth{0pt}
\tablecaption{Comparison of Apparent Magnitudes\label{tbl-3}}
\tablehead{
\colhead{Galaxy} &
\colhead{$\Delta m_J$} & 
\colhead{$\Delta m_K$} & 
\colhead{Reference} \\ 
\colhead{(1)} & 
\colhead{(2)} & 
\colhead{(3)} & 
\colhead{(4)}
}
\startdata
VCC 24   & -0.02 & -0.21 & 2MASS    \\
         &       & +0.04 & GOLDMine \\
VCC 144  & +0.09 & +0.27 & 2MASS    \\
VCC 213  & +0.24 & +0.13 & 2MASS    \\
VCC 324  & +0.25 & +0.19 & 2MASS    \\
         & +0.00 & -0.09 & GOLDMine \\
VCC 334  & -0.07 & -0.39 & GOLDMine \\
VCC 428  &       &       &          \\
VCC 459  & +0.33 & +0.04 & 2MASS    \\
         &       & +0.29 & GOLDMine \\ 
VCC 641  &       &       &          \\
VCC 802  &       & +0.56 & GOLDMine \\
VCC 848  & +1.06 & +1.00 & 2MASS    \\
         &       & +0.34 & GOLDMine \\
VCC 1313 &       &       &          \\
VCC 1374 & +0.39 & +0.44 & 2MASS    \\
         & -0.11 & +0.06 & GOLDMine \\
VCC 1437 & +0.09 & +0.10 & 2MASS    \\
VCC 1699 &       & +0.59 & GOLDMine \\
VCC 1725 & +0.40 & +0.52 & GOLDMine \\
VCC 2033 & +0.27 & +0.35 & 2MASS    \\
         &       & +0.58 & GOLDMine \\
\enddata
\tablecomments{
Col.~(1): Galaxy name; 
Col.~(2): Difference in $J$ magnitude (reference minus this paper); 
Col.~(3): Difference in $K_s$ magnitude (reference minus this paper); 
Col.~(4): Reference. 
}
\end{deluxetable}

\clearpage
\begin{deluxetable}{lcccccccccc}
\tablewidth{0pt}
\tabletypesize{\scriptsize}
\tablecaption{Photometric Parameters \label{tbl-4}}
\tablehead{
\colhead{Galaxy} &
\colhead{Filter} & 
\colhead{$m_I$} & 
\colhead{$m_S$} & 
\colhead{$m_G$} & 
\colhead{$m_T$} & 
\colhead{$\mu_{0S}$} & 
\colhead{$r_{0S}$} & 
\colhead{$\mu_{0G}$} & 
\colhead{$r_{0G}$} & 
\colhead{$r_{23(22)}$} \\
\colhead{} &
\colhead{} &
\colhead{(mag)} &
\colhead{(mag)} &
\colhead{(mag)} &
\colhead{(mag)} &
\colhead{(mag/sq.\arcsec)} &
\colhead{(\arcsec)} &
\colhead{(mag/sq.\arcsec)} &
\colhead{(\arcsec)} &
\colhead{(\arcsec)} \\
\colhead{(1)} &
\colhead{(2)} &
\colhead{(3)} &
\colhead{(4)} &
\colhead{(5)} &
\colhead{(6)} &
\colhead{(7)} &
\colhead{(8)} &
\colhead{(9)} &
\colhead{(10)} & 
\colhead{(11)}
}
\startdata
VCC 24   &   $J$ & 13.56 & 13.70 & 16.08 & 13.59 & 18.34 &  3.1 & 18.86 &  1.8 & 15.4 \\
         & $K_s$ & 12.78 & 13.09 & 14.58 & 12.84 & 17.65 &  3.0 & 17.73 &  2.1 & 14.0 \\
VCC 144  &   $J$ & 13.46 & 13.85 & 14.95 & 13.51 & 18.39 &  2.8 & 17.88 &  1.8 & 14.1 \\
         & $K_s$ & 12.57 & 12.89 & 14.18 & 12.60 & 17.28 &  2.7 & 17.25 &  2.0 & 13.4 \\
VCC 213  &   $J$ & 11.90 & 11.98 & 14.84 & 11.90 & 17.75 &  4.6 & 17.68 &  1.6 & 25.2 \\
         & $K_s$ & 11.12 & 11.21 & 14.07 & 11.14 & 16.94 &  4.5 & 16.85 &  1.6 & 23.9 \\
VCC 324  &   $J$ & 12.53 & 12.52 & 15.85 & 12.47 & 19.54 &  8.6 & 19.69 &  2.7 & 33.4 \\
         & $K_s$ & 11.83 & 11.92 & 14.66 & 11.84 & 18.86 &  8.3 & 18.60 &  2.8 & 29.7 \\
VCC 334  &   $J$ & 13.71 & 13.89 & 15.52 & 13.68 & 19.46 &  3.8 & 19.37 &  2.3 & 15.1 \\
         & $K_s$ & 13.13 & 13.25 & 15.25 & 13.09 & 18.77 &  3.7 & 18.85 &  2.1 & 13.7 \\
VCC 428  &   $J$ & 15.88 & 15.82 &   (a) & 15.82 & 21.05 &  5.2 &   (a) &  (a) & 12.9 \\
         & $K_s$ & 14.90 & 14.85 &   (a) & 14.85 & 20.39 &  6.0 &   (a) &  (a) & 12.9 \\
VCC 459  &   $J$ & 12.98 & 13.12 & 15.50 & 13.00 & 18.81 &  5.2 & 19.43 &  3.1 & 23.8 \\
         & $K_s$ & 12.31 & 12.46 & 14.81 & 12.34 & 18.10 &  5.1 & 18.76 &  3.2 & 21.8 \\
VCC 641  &   $J$ & 14.36 & 14.53 & 16.28 & 14.33 & 19.90 &  4.5 & 20.26 &  3.2 & 16.0 \\
         & $K_s$ & 13.27 & 13.37 & 16.00 & 13.28 & 18.70 &  4.4 & 20.62 &  4.3 & 16.5 \\
VCC 802  &   $J$ & 14.94 & 14.96 & 17.00 & 14.81 & 21.63 & 10.7 & 22.70 &  9.3 & 20.8 \\
         & $K_s$ & 14.25 & 14.23 & 18.55 & 14.21 & 20.57 &  9.2 & 20.35 &  1.5 & 18.4 \\
VCC 848  &   $J$ & 13.39 & 13.55 & 15.06 & 13.31 & 20.80 & 14.0 & 20.16 &  7.1 & 38.1 \\
         & $K_s$ & 12.57 & 12.71 & 14.02 & 12.42 & 20.17 & 15.5 & 19.06 &  6.8 & 36.7 \\
VCC 1313 &   $J$ & 15.82 & 16.59 & 16.55 & 15.82 & 20.91 &  2.8 & 19.89 &  2.4 &  7.2 \\
         & $K_s$ & 15.10 & 15.56 & 16.21 & 15.09 & 19.85 &  2.7 & 19.57 &  2.4 &  7.3 \\
VCC 1374 &   $J$ & 13.34 & 13.32 &   (a) & 13.32 & 19.88 &  9.5 &   (a) &  (a) & 34.0 \\
         & $K_s$ & 12.45 & 12.46 &   (a) & 12.46 & 18.95 &  9.2 &   (a) &  (a) & 32.3 \\
VCC 1437 &   $J$ & 12.84 & 13.14 & 14.53 & 12.87 & 18.28 &  3.7 & 17.21 &  1.6 & 18.4 \\
         & $K_s$ & 11.93 & 12.25 & 13.57 & 11.97 & 17.27 &  3.4 & 16.38 &  1.7 & 17.4 \\
VCC 1699 &   $J$ & 12.49 & 12.50 &   (a) & 12.50 & 19.86 & 12.3 &   (a) &  (a) & 44.2 \\
         & $K_s$ & 11.60 & 11.72 &   (a) & 11.72 & 19.04 & 12.1 &   (a) &  (a) & 41.4 \\
VCC 1725 &   $J$ & 12.73 & 12.81 &   (a) & 12.81 & 20.12 & 11.0 &   (a) &  (a) & 36.9 \\
         & $K_s$ & 11.74 & 11.81 &   (a) & 11.81 & 19.23 & 11.6 &   (a) &  (a) & 37.6 \\
VCC 2033 &   $J$ & 13.41 & 13.59 & 15.74 & 13.45 & 19.23 &  4.1 & 19.48 &  2.3 & 17.3 \\
         & $K_s$ & 12.45 & 12.71 & 15.00 & 12.59 & 18.34 &  4.1 & 18.58 &  2.2 & 16.8 \\
\enddata
\tablenotetext{(a)}{ Gaussian component not fitted }
\tablecomments{
Col.~(1): Galaxy name; 
Col.~(2): Filter; 
Col.~(3): Isophotal magnitude; 
Col.~(4): Sech magnitude (from sech law); 
Col.~(5): Gaussian magnitude (from Gaussian law); 
Col.~(6): Total (Sech + Gaussian) magnitude; 
Col.~(7): Central surface brightness of sech component; 
Col.~(8): Scale length of sech component; 
Col.~(9): Central surface brightness of Gaussian component; 
Col.~(10): Scale length of Gaussian component; 
Col.~(11): Semimajor axis $r_{23}$ corresponding to the $J=23$ mag arcsec$^{-2}$; semimajor axis $r_{22}$ corresponding to $K_s=22$ mag arcsec$^{-2}$. 
}
\end{deluxetable}

\clearpage
\begin{deluxetable}{lcccccccccc}
\tablewidth{0pt}
\tabletypesize{\scriptsize}
\tablecaption{Local Volume dIs with BCD-like profiles \label{tbl-5}}
\tablehead{
\colhead{Galaxy} &
\colhead{Filter} & 
\colhead{$m_I$} & 
\colhead{$m_S$} & 
\colhead{$m_G$} & 
\colhead{$m_T$} & 
\colhead{$\mu_{0S}$} & 
\colhead{$r_{0S}$} & 
\colhead{$\mu_{0G}$} & 
\colhead{$r_{0G}$} & 
\colhead{$r_{23(22)}$} \\
\colhead{} &
\colhead{} &
\colhead{(mag)} &
\colhead{(mag)} &
\colhead{(mag)} &
\colhead{(mag)} &
\colhead{(mag/sq.\arcsec)} &
\colhead{(\arcsec)} &
\colhead{(mag/sq.\arcsec)} &
\colhead{(\arcsec)} &
\colhead{(\arcsec)} \\
\colhead{(1)} &
\colhead{(2)} &
\colhead{(3)} &
\colhead{(4)} &
\colhead{(5)} &
\colhead{(6)} &
\colhead{(7)} &
\colhead{(8)} &
\colhead{(9)} &
\colhead{(10)} & 
\colhead{(11)}
}
\startdata
NGC 1569 &   $J$ &  9.06 &  9.38 & 10.67 &  9.09 & 17.99 & 21.0 & 17.14 & 10.6 & 111.3 \\
         & $K_s$ &  8.12 &  8.47 &  9.60 &  8.14 & 17.13 & 21.5 & 16.16 & 11.0 & 111.1 \\
NGC 3738 &   $J$ & 10.46 & 10.67 & 11.97 & 10.39 & 19.67 & 22.2 & 19.40 & 14.6 &  63.9 \\ 
         & $K_s$ &  9.61 &  9.72 & 11.40 &  9.51 & 18.67 & 21.7 & 18.45 & 12.2 &  81.7 \\ 
\enddata
\tablecomments{
Col.~(1): Galaxy name; 
Col.~(2): Filter; 
Col.~(3): Isophotal magnitude; 
Col.~(4): Sech magnitude (from sech law); 
Col.~(5): Gaussian magnitude (from Gaussian law); 
Col.~(6): Total (Sech + Gaussian) magnitude; 
Col.~(7): Central surface brightness of sech component; 
Col.~(8): Scale length of sech component; 
Col.~(9): Central surface brightness of Gaussian component; 
Col.~(10): Scale length of Gaussian component; 
Col.~(11): Semimajor axis $r_{23}$ corresponding to the $J=23$ mag arcsec$^{-2}$; semimajor axis $r_{22}$ corresponding to $K_s=22$ mag arcsec$^{-2}$. 
}
\end{deluxetable}

\clearpage
\begin{deluxetable}{lcccccccccc}
\tablewidth{0pt}
\tabletypesize{\scriptsize}
\tablecaption{ Sech and Sech+Gaussian SBP fitting for six BCDs with dI-like profiles \label{tbl-6}}
\tablehead{
\colhead{Galaxy} &
\colhead{Filter} &
\colhead{Fit} &
\colhead{rms} &
\colhead{$m_I$} & 
\colhead{$m_T$} & 
\colhead{$\mu_{0S}$} & 
\colhead{$r_{0S}$} & 
\colhead{$\mu_{0G}$} & 
\colhead{$r_{0G}$} & 
\colhead{$r_{23(22)}$} \\
\colhead{} &
\colhead{} &
\colhead{} &
\colhead{} &
\colhead{(mag)} &
\colhead{(mag)} &
\colhead{(mag/sq.\arcsec)} &
\colhead{(\arcsec)} &
\colhead{(mag/sq.\arcsec)} &
\colhead{(\arcsec)} &
\colhead{(\arcsec)} \\
\colhead{(1)} &
\colhead{(2)} &
\colhead{(3)} &
\colhead{(4)} &
\colhead{(5)} &
\colhead{(6)} &
\colhead{(7)} &
\colhead{(8)} &
\colhead{(9)} &
\colhead{(10)} & 
\colhead{(11)}
}
\startdata
VCC 428  &   $J$ &   S & 0.010 & 15.88 & 15.82 & 21.05 &  5.2 &    -- &   -- & 12.9 \\
         &   $J$ & S+G & 0.013 & 15.88 & 15.76 & 21.90 &  6.8 & 21.69 &  4.9 & 11.4 \\
         & $K_s$ &   S & 0.041 & 14.90 & 14.85 & 20.39 &  6.0 &    -- &   -- & 12.9 \\
         & $K_s$ & S+G & 0.044 & 14.90 & 14.77 & 20.49 &  6.5 & 20.47 &  0.9 & 13.3 \\
	 
VCC 848  &   $J$ &   S & 0.037 & 13.39 & 13.61 & 19.69 &  8.2 &    -- &   -- & 23.1 \\
         &   $J$ & S+G & 0.009 & 13.39 & 13.31 & 20.80 & 14.0 & 20.16 &  7.1 & 38.1 \\
         & $K_s$ &   S & 0.055 & 12.57 & 12.82 & 18.76 &  7.7 &    -- &   -- & 28.3 \\
         & $K_s$ & S+G & 0.020 & 12.57 & 12.42 & 20.17 & 15.5 & 19.06 &  6.8 & 36.7 \\
	 
VCC 1313 &   $J$ &   S & 0.017 & 15.82 & 15.80 & 19.43 &  2.0 &    -- &   -- &  6.2 \\
         &   $J$ & S+G & 0.007 & 15.82 & 15.82 & 20.91 &  2.8 & 19.89 &  2.4 &  7.2 \\
         & $K_s$ &   S & 0.028 & 15.10 & 15.10 & 18.82 &  2.1 &    -- &   -- &  7.6 \\
         & $K_s$ & S+G & 0.027 & 15.10 & 15.09 & 19.85 &  2.7 & 19.57 &  2.4 &  7.3 \\
	 
VCC 1374 &   $J$ &   S & 0.033 & 13.34 & 13.32 & 19.88 &  9.5 &    -- &   -- & 34.0 \\
         &   $J$ & S+G & 0.033 & 13.34 & 13.42 & 19.04 &  7.2 &   n/a &  6.9 & 31.3 \\
         & $K_s$ &   S & 0.043 & 12.45 & 12.46 & 18.95 &  9.2 &    -- &   -- & 32.3 \\
         & $K_s$ & S+G & 0.074 & 12.45 & 12.55 & 17.96 &  6.8 &   n/a &  6.8 & 30.2 \\
	 
VCC 1699 &   $J$ &   S & 0.017 & 12.49 & 12.50 & 19.86 & 12.3 &    -- &   -- & 44.2 \\
         &   $J$ & S+G & 0.016 & 12.49 & 12.54 & 19.77 & 11.7 &   n/a &  6.4 & 43.0 \\
         & $K_s$ &   S & 0.036 & 11.60 & 11.72 & 19.04 & 12.1 &    -- &   -- & 41.4 \\
         & $K_s$ & S+G & 0.022 & 11.60 & 11.49 & 20.36 & 21.0 & 19.42 & 11.4 & 46.0 \\
	 
VCC 1725 &   $J$ &   S & 0.025 & 12.73 & 12.81 & 20.12 & 11.0 &    -- &   -- & 36.9 \\
         &   $J$ & S+G & 0.027 & 12.73 & 12.73 & 20.30 & 12.2 & 21.59 &  5.1 & 38.9 \\
         & $K_s$ &   S & 0.039 & 11.74 & 11.81 & 19.23 & 11.6 &    -- &   -- & 37.6 \\
         & $K_s$ & S+G & 0.049 & 11.74 & 11.80 & 18.55 & 10.1 &   n/a & 11.6 & 39.2 \\
\enddata
\tablecomments{
Col.~(1): Galaxy name; 
Col.~(2): Filter; 
Col.~(3): Fit (S=sech; S+G=sech+Gaussian); 
Col.~(4): rms error of fit; 
Col.~(5): Isophotal magnitude; 
Col.~(6): Total magnitude (sech or sech+Gaussian); 
Col.~(7): Central surface brightness of sech component; 
Col.~(8): Scale length of sech component; 
Col.~(9): Central surface brightness of Gaussian component; 
Col.~(10): Scale length of Gaussian component; 
Col.~(11): Semimajor axis $r_{23}$ corresponding to the $J=23$ mag arcsec$^{-2}$, semimajor axis $r_{22}$ corresponding to $K_s=22$ mag arcsec$^{-2}$. 
}
\end{deluxetable}

\clearpage

\end{document}